\documentclass[10pt]{wlscirep}
\usepackage{amsmath}
\usepackage{bm}
\usepackage[utf8]{inputenc}
\usepackage[T1]{fontenc}
\usepackage{caption}
\usepackage{subfigure}
\usepackage{graphicx}
\usepackage{epstopdf}
\title{Identifying vital nodes based on reverse greedy method}

\author[1,*]{Tao Ren}
\author[1]{Zhe Li}
\author[1]{Yi Qi}
\author[1]{Yixin Zhang}
\author[1]{Simiao Liu}
\author[1]{Yanjie Xu}
\author[2,*]{Tao Zhou}
\affil[1]{Software College, Northeastern University of China, Shenyang, 110819, P. R. China}
\affil[2]{CompleX Lab, University of Electronic Science and Technology of China, Chengdu, 611731, P. R. China}

\affil[*]{chinarentao@163.com, zhutou@ustc.edu}

\begin{abstract}
The identification of vital nodes that maintain the network connectivity is a long-standing challenge in network science. In this paper, we propose a so-called reverse greedy method where the least important nodes are preferentially chosen to make the size of the largest component in the corresponding induced subgraph as small as possible. Accordingly, the nodes being chosen later are more important in maintaining the connectivity. Empirical analyses on ten real networks show that the reverse greedy method performs remarkably better than well-known state-of-the-art methods.
\end{abstract}
\begin{document}
\flushbottom
\maketitle
\thispagestyle{empty}
\section*{Introduction}
Network science is playing an increasingly significant role in many domains including physics, sociology, engineering, biology, management, and so on \cite{1}. Because of the heterogeneous nature of real networks \cite{2}, the overall connectivity of complex networks may depend on a small set of nodes, usually named as hub nodes. Taking the Internet as an example, several vital nodes attacked deliberately may lead to the collapse of the whole network \cite{3}. Therefore, an efficient algorithm to identify vital nodes that have critical impacts on the network connectivity can help to better prevent catastrophic outages in power grids or the Internet \cite{3,4,5,6}, maintain the connectivity or design efficient attacking strategies for communication networks \cite{7}, improve urban transportation capacity with low cost \cite{8}, enhance robustness of financial networks \cite{9}, and so on.

Till far, to identify vital nodes for network connectivity, the majority of known methods only make use of the structural information \cite{10}. Typical representatives include degree centrality \cite{11} (DC), H-index \cite{12}, k-shell decomposition method \cite{13} (KS), PageRank \cite{14} (PR), LeaderRank \cite{14.5}, closeness centrality \cite{15} (CC), betweenness centrality \cite{16} (BC), and so on. For DC, nodes with larger degrees are more vital. For H-index, nodes connecting with many large-degree neighbors are more important. KS assigns a k-shell index to each node based on its topological location, where nodes closer to the core of the network will get higher k-shell indices, and nodes in the periphery will get lower k-shell indices. The nodes with higher k-shell indices are considered to be more vital. PR suggests that the importance of a node is determined by the influences of its neighbors. CC claims that a node averagely closer to other nodes is more vital while BC assumes that a node locating in many shortest paths is of high importance. Recently, Morone and Makse \cite{17} proposed a novel index called collective influence (CI), which is based on the site percolation theory and can find out the minimal set of nodes that are crucial for the global connectivity. CI performs remarkably better than many previous methods in identifying the nodes' importance for network connectivity \cite{17,17.5}.

This paper proposed a novel method named reverse greedy (RG) method. The first word stands for the process that we add nodes one by one to an empty network, which is inverse to the usual process that removes nodes from the original network. The second word emphasizes that we choose the nodes added by minimizing the size of the largest component. Empirical analyses on ten real networks show that RG performs remarkably better than well-known state-of-the-art methods.

\section*{Results}
\subsection*{Algorithms}
\begin{figure}[htbp]
\centering
\includegraphics[width=12cm]{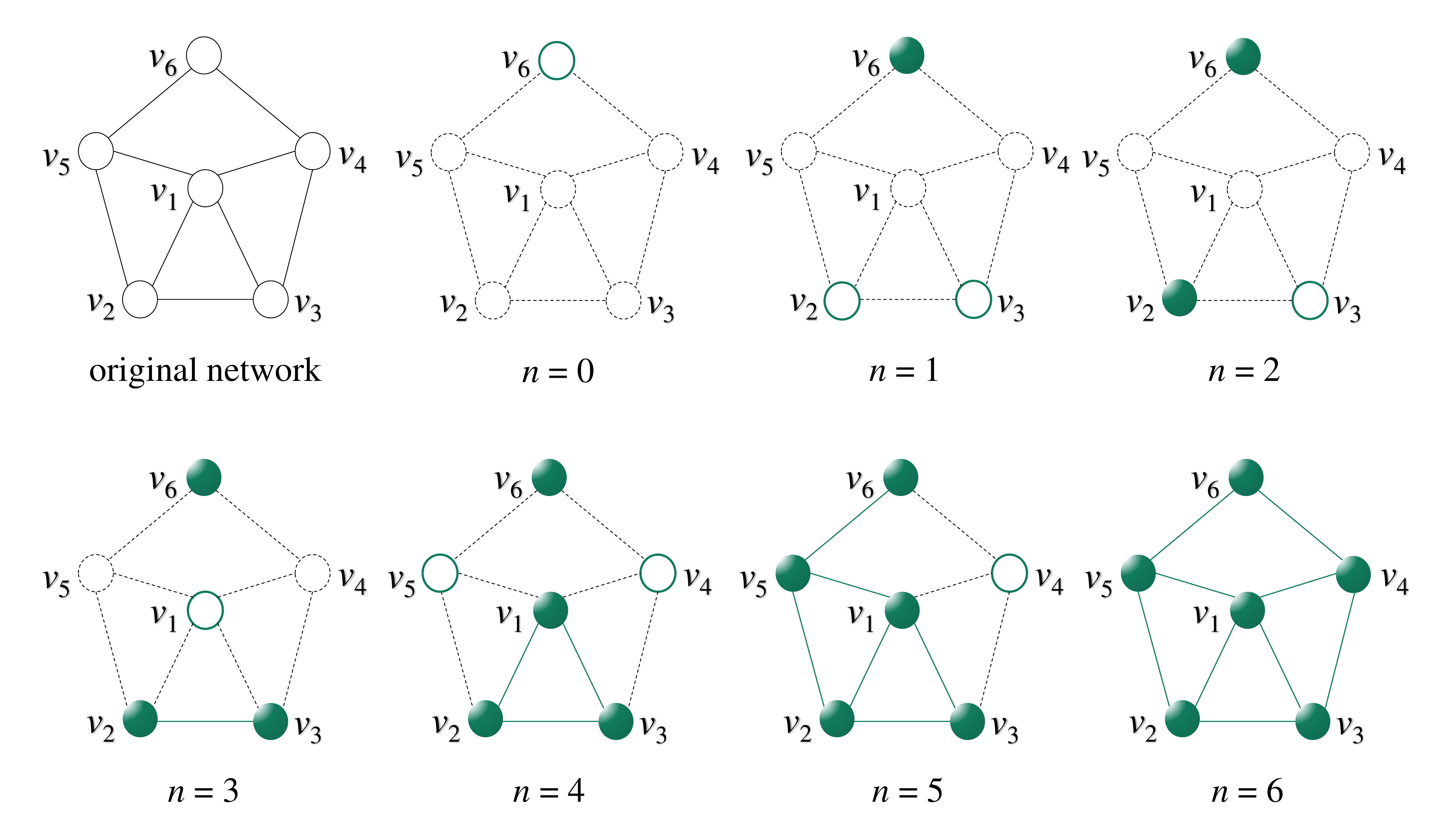}
\caption{The process of RG in a network with six nodes. Here we use degree as the feature $f$ and for convenience in the later description we set $\epsilon=0.01$ (note that, we only require that $\epsilon$ is positive yet small enough). Initially, the network is empty. At the first time step, $G_1^{\max}(v_{1})=G_1^{\max}(v_{2})=G_1^{\max}(v_{3})=G_1^{\max}(v_{4})=G_1^{\max}(v_{5})=G_1^{\max}(v_{6})=1$, and $cost(v_{1},1)=1.04$, $cost(v_{2},1)=1.03$, $cost(v_{3},1)=1.03$, $cost(v_{4},1)=1.03$, $cost(v_{5},1)=1.03$, $cost(v_{6},1)=1.02$. Therefore, we add node $v_{6}$ into the network because $cost(v_{6},1)$ is the smallest. In the second time step, $G_2^{\max}(v_{1})=G_2^{\max}(v_{2})=G_2^{\max}(v_{3})=1$, and $G_2^{\max}(v_{4})=G_2^{\max}(v_{5})=2$, so we only compare three candidates $v_1$, $v_2$ and $v_3$. Since $cost(v_{1},1)=1.04$, $cost(v_{2},1)=1.03$ and $cost(v_{3},1)=1.03$, we randomly select a node from $\{v_2,v_3\}$. Here we choose $v_2$ for example. Repeat this process until all nodes are added into the network. Finally, we get the ranking of nodes as $\{v_{4}, v_{5}, v_{1}, v_{3}, v_{2}, v_{6}\}$, in an inverse order of the additions. The symbol $n$ in the bottom of each plot stands for the corresponding time step.}
\label{fig:1}
\end{figure}

The core of the RG algorithm is the reverse process, which adds nodes one by one to an empty network while minimizes the cost function until all nodes in the considered network are added. Then, nodes are ranked inverse to the order of additions, that is to say, the later added nodes are more important in maintaining the network connectivity. Denote $G(V,E)$ the original network under consideration, where $V$ and $E$ are the sets of nodes and edges, respectively. This paper focuses on simple networks, where the weights and directions of edges are ignored, and the self loops are not allowed. The reverse process starts from an empty network $G_0(V_0,E_0)$, where $V_0=\emptyset$ and $E_0=\emptyset$. At the $(n+1)$th time step, one node from the remaining set $V-V_n$ is selected to add into the current network $G_n(V_n,E_n)$ to form a new network of $(n+1)$ nodes, say $G_{n+1}(V_{n+1},E_{n+1})$. Note that, all progressive networks $G_n$ ($n=0,1,2,\cdots,N$, with $N$ being the size of the original network $G$) in the process are induced subgraphs of $G$. For example, $G_n$ is consisted of all edges in $G$ with both two ends belonging to $V_n$. According to the greedy strategy, the selected node $i$ should minimize the size of the largest component in $G_{n+1}$. If there are multiple nodes satisfying this condition, we will choose the one with the help of another structural feature of the node $i$ in $G$ (e.g., degree, betweenness, and so on). Therefore, the cost function can be defined as
\begin{equation}\label{1}
cost(i,n+1)=G^{\max}_{n+1}(i)+\epsilon f(i),
\end{equation}
where $G^{\max}_{n+1}(i)$ is the size of the largest component after adding node $i$ into $G_n$, $f(i)$ is a certain structural feature of node $i$ in $G$, and $\epsilon$ is a very small positive parameter that works only when $G^{\max}_{n+1}(\bullet)$ are indistinguishable for multiple nodes. Each time step, we add the node minimizing the cost function into the network, and if there are still multiple nodes with the minimum cost, we will select one of them randomly. This process stops after $N$ time steps, namely all nodes are added with $G_N\equiv G$. An illustration of such process in a small network is shown in Figure \ref{fig:1}.

\subsection*{Data Description}
In this paper, ten real networks from disparate fields are used to test the performance of RG, including two collaboration networks (Jazz and NS), one communication network (Email), three social networks (PB, Sex and Facebook), one transportation network (USAir), one infrastructure network (Power), one technological network (Router) and one citation network (HepPh). Jazz \cite{18} is a collaboration network of jazz musicians. NS \cite{19} is a co-authorship network of scientists working on network science. Email \cite{20} describes email interchanges between users including faculty, researchers, technicians, managers, administrators, and graduate students of the Rovira i Virgili University. PB \cite{21} is a network of US political blogs. Sex \cite{22} is a bipartite network in which nodes are females (sex sellers) and males (sex buyers) and edges between them are established when males write posts indicating sexual encounters with females. Facebook \cite{23} is a sample of the friendship network of Facebook users. USAir \cite{24} is the US air transportation network. Power \cite{25} is the power grid of the western United States. Router \cite{26} is a symmetrized snapshot of the structure of the Internet at the level of autonomous systems. HepPh \cite{27} is a citation network of high energy physics phenomenology. These networks' topological features (including the number of nodes, the number of edges, the average degree, the clustering coefficient \cite{25}, the assortative coefficient \cite{28} and the degree heterogeneity \cite{29}) are shown in Table \ref{tab:1}.
\begin{table}[htbp]
\centering
\begin{tabular}{ccccccccc}
\hline
\textbf{Networks} & \bm{$N$} & \bm{$E$}  &   \bm{$\langle k\rangle$}  & \bm{$C$} & \bm{$r$} & \bm{$H$} \\
\hline
Jazz&198&2742&27.6970&0.6334&0.0202&1.3951\\
NS&379&914&4.8232&0.7981&-0.0817&1.6630\\
Email&1133&5451&9.6222&0.2540&0.0782&1.9421\\
PB&1222&16714&27.3552&0.3600&-0.2213&2.9707\\
Sex&15810&38540&4.8754&0&-0.1145&5.8276\\
Facebook&63731&817090&25.6418&0.2532&0.1769&3.4331\\
USAir&332&2126&12.8072&0.7494&-0.2079&3.4639\\
Power&4941&6594&2.6691&0.1065&0.0035&1.4504\\
Router&5022&6258&2.4922&0.0329&-0.1384&5.5031\\
HepPh&34546&420877&24.3662&0.2962&-0.0063&2.6055\\
\hline
\end{tabular}
\caption{\label{tab:1}The basic topological features of the ten real networks. $N$ and $E$ are the number of nodes and edges, $\left \langle k \right \rangle$ is the average degree, $C$ is the clustering coefficient, $r$ is the assortative coefficient and $H$ is the degree heterogeneity.}
\end{table}

\subsection*{Empirical Results}
We apply the widely used metric called \emph{robustness} $R$ \cite{30} to evaluate algorithms' performance. Given a network, we remove one node at each time step and calculate the size of the largest component of the remaining network until the remaining network is empty. The \emph{robustness} $R$ is defined as \cite{30}
\begin{equation}\label{2}
R=\frac{1}{N}\sum_{Q=1}^{N}S(Q),
\end{equation}
where $S(Q)$ is the number of nodes in the largest component divided by $N$ after removing $Q$ nodes. The normalization factor $1/N$ ensures that the values of $R$ of networks with different sizes can be compared. Obviously, a smaller $R$ means a quicker collapse and thus a better performance.

\begin{figure}[htbp]
\centering
\includegraphics[width=10cm]{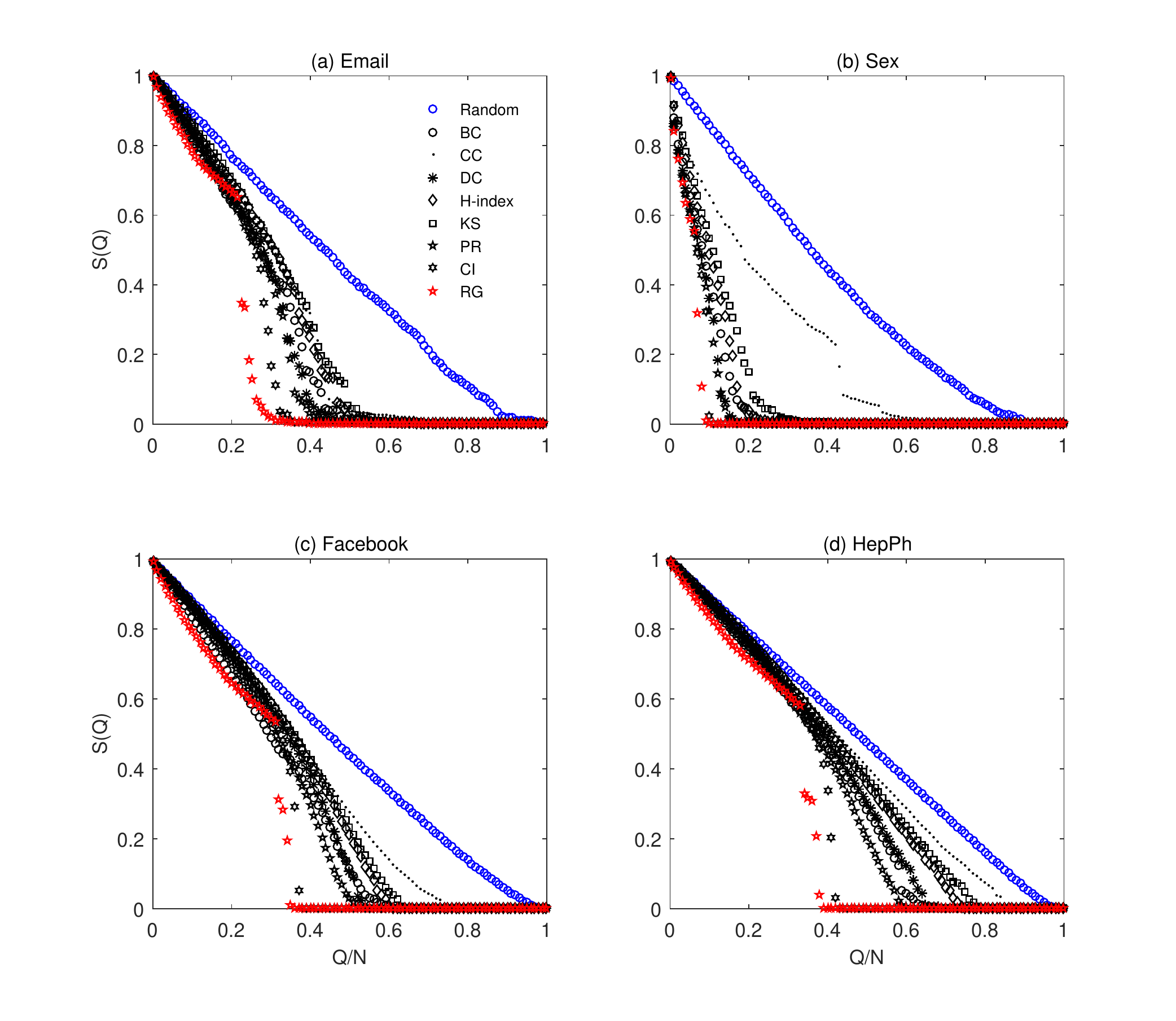}
\caption{Comparing the performance of the background benchmark (random removal, denoted by blue circles), RG (red stars) and the other seven benchmark algorithms (black symbols). The $X$-axis is the fraction of nodes being removed (i.e., $Q/N$), and the $Y$-axis denotes the number of nodes in the largest component divided by $N$ (i.e., $S(Q)$). The four selected networks are (a) Email, (b) Sex, (c) Facebook, and (d) HepPh, respectively. Other networks exhibit similar results.}
\label{fig:2}
\end{figure}

Figure \ref{fig:2} shows the collapsing processes of four representative networks, resulted from the node removal by RG and other benchmark algorithms (see details about these benchmark algorithms in Methods). Obviously, RG can lead to much faster collapse than all other algorithms, and CI is the second best algorithm. Table \ref{tab:2} compares the \emph{robustness} $R$ of RG and other benchmarks. As shown in Table \ref{tab:2}, every algorithm is better than the random removal and to our surprise, for all the ten networks, RG is always the best. In most cases, CI is the second best algorithm. One can further observe that the advantage of RG is particularly significant for sparse networks, such as Power and Router.

\begin{table}[htbp]
  \centering
    \begin{tabular}{cccccccccc}
    \hline
    \textbf{Networks}&\textbf{Random}&\textbf{BC}&\textbf{CC}&\textbf{DC}&\textbf{H-index}&\textbf{KS}&\textbf{PR}&\textbf{CI}&\textbf{RG}\\
    \hline
    Jazz    &0.4808&0.3956&0.4199&0.4409&0.4497&0.4571&0.4262&0.3913&\textbf{0.3477}\\
    NS      &0.2752&0.0488&0.1336&0.0540&0.1155&0.1582&0.0524&0.0551&\textbf{0.0252}\\
    Email   &0.4442&0.2578&0.2893&0.2519&0.2836&0.2937&0.2395&0.2231&\textbf{0.1844}\\
    PB      &0.4615&0.2192&0.2908&0.2286&0.2578&0.2611&0.2155&0.1968&\textbf{0.1740}\\
    Sex     &0.3842&0.0841&0.2208&0.0725&0.0981&0.1142&0.0690&0.0604&\textbf{0.0513}\\
    Facebook&0.4545&0.2935&0.3570&0.3137&0.3328&0.3389&0.2893&0.2671&\textbf{0.2372}\\
    USAir   &0.4321&0.1129&0.1442&0.1228&0.1498&0.1588&0.1072&0.1105&\textbf{0.0942}\\
    Power   &0.2069&0.0656&0.1973&0.0634&0.1090&0.2628&0.0594&0.0489&\textbf{0.0088}\\
    Router  &0.3044&0.0142&0.0686&0.0121&0.0136&0.0276&0.0136&0.0140&\textbf{0.0063}\\
    HepPh   &0.4765&0.3504&0.4259&0.3664&0.3931&0.4022&0.3371&0.3015&\textbf{0.2657}\\
    \hline
    \end{tabular}%
  \caption{\label{tab:2}The performance, measured by \emph{robustness} $R$, of the eight ranking methods on ten real networks. The best performed method for each network, namely the lowest $R$ in the corresponding row, is emphasized in bold. Notice that, we use the random removal (Random) as the background benchmark in order to show the improvement by each method. The radius $\ell$ in CI is set to 2, and the feature $f(i)$ in RG is the degree of node $i$.}
\end{table}%

\section*{Discussion}
To our knowledge, most previous methods directly identify the critical nodes by looking at the effects due to their removal \cite{10}. In contrast, our method tries to find out the least important nodes, so that the remaining ones are those critical nodes. To our surprise, such a simple idea eventually results in an efficient algorithm that outperforms many well-known benchmark algorithms. Beyond the percolation process considered in this paper, the reverse method provides a novel angle of view that may find successful applications in some other network-based optimization problems related to certain rankings of nodes or edges.

Lastly, we would like to emphasize that the current version of the RG algorithm is just the simplest implementation of the above reverse idea. For example, instead of degree, $f(i)$ can be designed in a sophisticated way to improve the algorithm’s performance. In addition, the simple adoption of the greedy strategy may bring us to some local optimums. Such shortage can be to some extent overcame by introducing the beam search \cite{31}, which searches for the best set of $m$ nodes adding to the network that optimizes the cost function. The present algorithm is the special case for $m=1$. Although beam search is still a kind of greedy strategy, it usually performs much better when $m$ is sufficiently large. At the same time, the beam search with large $m$ costs a lot on time and space. Therefore, how to find a good tradeoff is also an open challenge in real practice.

\section*{Methods}
\subsection*{Benchmark Centralities}
Degree Centrality \cite{11} of node $i$ is defined as
\begin{equation}\label{5}
DC(i)=\sum_j a_{ij},
\end{equation}
where $A=\{a_{ij}\}$ is the adjacency matrix, that is, $a_{ij}$ = 1 if $i$ and $j$ are directly connected and 0 otherwise.

H-index \cite{12} of node $i$, denoted by $H(i)$, is defined as the maximal integer satisfying that there are at least $H(i)$ neighbors of node $i$ whose degrees are all no less than $H(i)$. Such index is an extension of the famous H-index in scientific evaluation \cite{32} to network analysis.

PageRank \cite{14} of node $i$ is defined as the solution of the equations
\begin{equation}\label{6}
PR_{i}(t)=s\sum_{j=1}^{N}a_{ji}\frac{PR_{j}(t-1)}{k_{j}}+(1-s)\frac{1}{N},
\end{equation}
where $k_{j}$ is the degree of node $j$ and $s$ is a free parameter controlling the probability of a random jump. In this paper, $s$ is set to $0.85$.

Closeness Centrality \cite{15} of node $i$ is defined as
\begin{equation}\label{7}
CC(i)=\frac{N-1}{\sum\limits_{j\neq i} d_{ij}},
\end{equation}
where $d_{ij}$ is the shortest distance between nodes $i$ and $j$.

Betweenness Centrality \cite{16} of node $i$ is defined as
\begin{equation}\label{8}
BC(i)=\sum_{s\ne{i},s\ne{t},i\ne{t}}\frac{g_{st}(i)}{g_{st}},
\end{equation}
where $g_{st}$ is the number of shortest paths between nodes $s$ and $t$, and $g_{st}(i)$ is the number of shortest paths between nodes $s$ and $t$ that pass through node $i$.

Collective Influence \cite{17} (CI) of node $i$ is defined as
\begin{equation}\label{9}
CI(i)=(k_{i}-1)\sum_{j\in \partial ball(i,\ell)}(k_{j}-1),
\end{equation}
where $ball(i,\ell)$ is the set of nodes inside a ball of radius $\ell$, consisted of all nodes with distances no more than $\ell$ from node $i$, and $\partial ball(i,\ell)$ is the frontier of this ball.

\section*{Data Availability}
All relevant data are available at https://github.com/MLIF/Network-Data2.

\section*{Acknowledgements}
The authors acknowledge DataCastle to hold the related world-wide competition and to share the data. This work is partially supported by National Natural Science Foundation of China (61473073, 61104074, 61433014), Fundamental Research Funds for the Central Universities (N161702001, N171706003, N181706001, N182608003).
\section*{Author Contributions}
T.R., Y.Q., Z.L. and T.Z. devised the research project. Y.Q., Y.X.Z. and S.M.L. performed the research. T.R., Z.L., Y.Q., Y.X.Z., S.M.L. and T.Z. analyzed the data. T.R., Z.L., Y.X.Z., S.M.L., Y.J.X. and T.Z. wrote the paper.
\section*{Additional Information}
Competing Interests: The authors declare no competing interests.
\end{document}